\documentclass[letterpaper, 12 pt]{article}

\usepackage{graphicx}                                     
\usepackage{amsmath,amssymb}
\usepackage{mathrsfs}
\usepackage{multicol}

\title{\Huge\bf
A Generalized Robust Filtering Framework for Nonlinear
Differential-Algebraic Systems}

\date{ }

\author{Masoud Abbaszadeh
\thanks{{\tt\small e-mail: masoud@ualberta.net}}%
}

\begin{document}

\maketitle \begin{center}\thanks{Department of
        Research and Development, Maplesoft, Waterloo, Ontario, Canada}\end{center}

\begin{abstract}
A generalized dynamical robust nonlinear filtering
framework is established for a class of Lipschitz differential algebraic systems, in
which the nonlinearities appear both in the state and measured
output equations. The system is assumed to be affected by norm-bounded disturbance
and to have both norm-bounded uncertainties in the
realization matrices as well as nonlinear model
uncertainties. We synthesize a robust $H_{\infty}$ filter through
semidefinite programming and strict linear matrix inequalities (LMIs).
The admissible Lipschitz constants of the nonlinear functions
are maximized through LMI
optimization. The resulting $H_{\infty}$ filter guarantees
asymptotic stability of the estimation error dynamics with
prespecified disturbance attenuation level and is robust against
time-varying parametric uncertainties as well as Lipschitz nonlinear
additive uncertainty. Explicit bound on the tolerable nonlinear
uncertainty is derived based on a norm-wise robustness analysis.\\
\end{abstract}

\emph{keywords}: Robust Filtering, Nonlinear $H_{\infty}$,
Differential Algebraic Equations, Descriptor Systems,
Semidefinite Programming


\section{Introduction}
State estimation and filtering of nonlinear dynamical
systems has been a subject of extensive research in recent years
due to its theoretical and practical importance.
State estimators are essential for observer-based control, fault detection and isolation,
prediction and smoothing, and monitoring purposes.
Generalizing the state space modeling,
descriptor systems can characterize a larger class of
systems than conventional state space models
and can \emph{describe} the physics of the system more precisely.
Descriptor systems, also referred to as singular systems or differential-algebraic
equation (DAE) systems, arise from an inherent and natural modeling
approach, and have vast applications in engineering disciplines such as
power systems, network and circuit analysis, and multibody mechanical systems,
as well as in social and economic sciences.
Model based development (MBD) processes are
adopted in advanced control methodologies in areas such as in automotive,
energy, mechatronics and aerospace.
Recently, DAE systems have become a fundamental part of physical
modeling and simulation of dynamical system.
The number of models described by DAEs has been rapidly growing, partly due to modern modeling
tools, such as those based on the \emph{Modelica} object oriented modeling language \cite{Fritzson2004,Cellier2006}.

As more and more DAE models are available, it is
natural to directly use them for controller or filter design.
Many approaches have been developed to design state
observers for descriptor systems.
In \cite{Dai, Boutayeb1, Hou, Darouch1, Darouch2, Darouach.etal2008_1, Uezato, Lu_G3, He_Wang, Masubuchi,
Shields, Zimmer,Boulkroune.etal2010_1,Boulkroune.etal2010_2} various methods of observer design for linear and
nonlinear descriptor systems have been proposed. In \cite{Boutayeb1}
an observer design procedure is proposed for a class nonlinear
descriptor systems using an appropriate coordinate transformation.
In \cite{Shields}, the authors address the unknown input observer
design problem dividing the system into two dynamic and static
subsystems. References \cite{Darouach.etal2008_1, Lu_G3} study the full order and
reduced order observer design for Lipschitz nonlinear systems.

\paragraph{Three aspects of robust filtering approaches:}
A fundamental limitation encountered in conventional observer theory
is that it cannot guarantee observer performance in the presence of model uncertainties
and/or disturbances and measurement noise.
One of the most popular ways to deal with the nonlinear state estimation problem is the extended Kalman
filtering. However, the requirements of specific noise statistics and weakly nonlinear dynamics, has restricted its
applicability to nonlinear systems. To deal with the nonlinear filtering problem in
the presence of model uncertainties and unknown exogenous disturbances, we can resort the
\emph{robust} $H_{\infty}$ or similar filtering approaches.
A \emph{robust filtering} approach accomplishes the following objectives:
 \begin{itemize}
   \item \textbf{Stability:} In the absence of external disturbances, the filter error asymptotically converges to zero.
Moreover, our design is such that it
can maximize the size of the Lipschitz constant that can be tolerated in the system which directly translates
to the expansion of the admissible region of operation.
   \item \textbf{Robustness:} The design is robust with respect to uncertainties in the nonlinear plant
model.
   \item \textbf{Filtering:} The effect of exogenous disturbances on the filter error can be minimized.
 \end{itemize}

To deal with the nonlinear state observation problem in
the presence of model uncertainties and unknown exogenous disturbances, the
robust $H_{\infty}$ filtering was proposed as an effective approach.
In $H_{\infty}$ filtering, the $\mathcal{L}_{2}$ gain from the exogenous disturbance to the filter error is guaranteed
to be less than a prespecified level. Therefore, this $\mathcal{L}_{2}$
gain minimization is in fact an \emph{energy-to-energy} filtering problem.
The disturbance can be any signal with finite energy, either stochastic (with unknown statistics) or deterministic.
See for example references \cite{deSouza1}, \cite{deSouza2} and the references therein.
In these references, the authors consider a class of continuous-time
nonlinear system satisfying a Lipschitz continuity condition. The mathematical system model is
assumed to be affected by time-varying parametric uncertainties and norm bounded disturbances affect the measurements.
Under these conditions they obtain Riccati-based sufficient conditions for the stability of the
proposed filter with guaranteed disturbance attenuation level.
In the absence of disturbance, of course, the solution of the filtering problem renders an asymptotic observer
whose state converges to the plant state.
We point out here that the elegance of the Riccati approach comes at the price of somewhat restrictive
regularity assumptions required in the solution of the synthesis problem.
These restrictions are not inherent in the $H_{\infty}$ formulation but are a consequence of the
Riccati approach and can be relaxed using Linear Matrix inequalities (LMIs).

In this work, we study the robust nonlinear $H_{\infty}$ filtering criterion
for continuous-time Lipschitz DAE (descriptor) systems in the presence of
disturbance and model uncertainties, in a linear matrix inequalities (LMI) framework.
The linear matrix inequalities
proposed here are developed in such a way that the admissible
Lipschitz constant of the system is maximized through LMI
optimization. This adds the important extra feature to the filter,
making it robust against a class of nonlinear uncertainty. Securing the
same filter features, the LMI optimization approach to nonlinear
filtering for the conventional state space models can
be found in \cite{Abbaszadeh3,abbaszadeh2008lmi,abbaszadeh2012generalized}
and \cite{Abbaszadeh4, Abbaszadeh5}
in continuous and discrete time domains, respectively.
The results given here generalize our previous results in that: i) extend the
model from conventional state space to descriptor models, ii)
consider nonlinearities in both the state and output equations,
and iii) generalize the filter structure by
proposing a general dynamical filtering framework that can easily
capture both dynamic and static-gain filter structures. The
proposed dynamical structure has additional degree of freedom compared to
conventional static-gain filters and consequently is capable of
robustly stabilizing the filter error dynamics for
systems for which an static-gain filter cannot be found. Besides,
for the cases that both static-gain and dynamic filters exist, the
maximum admissible Lipschitz constant obtained using the proposed
dynamical filter structure can be much larger than that of the
static-gain filter. The result is a filter with a
prespecified disturbance attenuation level which guarantees
asymptotic stability of the estimation error dynamics and is robust
against Lipschitz nonlinear uncertainties as well as time-varying
parametric uncertainties, simultaneously.

In section II, the problem statement and some
preliminaries are mentioned. In section III, we propose a new method
for robust $H_{\infty}$ filter design for nonlinear descriptor
uncertain systems based on semidefinite programming (SDP).
In Section IV, the SDP problem of Section III is
converted into strict LMIs. Section V, is devoted to robustness
analysis in which an explicit bound on the tolerable nonlinear
uncertainty is derived. In section VI, we show the effectiveness of our proposed filter
design procedure through an illustrative example. Section VII
includes our concluding remarks and some proposed future research directions.


\section{Preliminaries and Problem Statement}

Consider the following class of continuous-time uncertain nonlinear
descriptor systems:
\begin{align}
\left(\Sigma_{s} \right): \mathbf{E}\dot{x}(t)&=(A+\Delta
A(t))x(t)+\Phi(x,u)+B w(t)\label{sys1}
\\ y(t)&=(C+\Delta C(t))x(t)+\Psi(x,u)+Dw(t)\label{sys2}
\end{align}
where $x\in {\mathbb R} ^{n} ,u\in {\mathbb R} ^{m} ,y\in {\mathbb
R}^{p} $ and $\Phi(x,u)$ and $\Psi(x,u)$ contain nonlinearities of
second order or higher. $\mathbf{E}$, $A$, $B$, $C$ and $D$ are
constant matrices with compatible dimensions. The matrix $\mathbf{E}$,
which following the analogy from the multibody dynamics modeling,
is often referred to as the \emph{mass matrix}, may be
singular. When the matrix $\mathbf{E}$ is singular,
the above form is equivalent to a set of
differential-algebraic equations (DAEs) \cite{Dai}.
In other words, the dynamics of descriptor systems, comprise a set of differential
equations together with a set of algebraic constraints. Unlike conventional
state space systems in which the initial conditions can be
freely chosen in the operating region, in the descriptor systems,
initial conditions must be \emph{consistent}, i.e.
they should satisfy the algebraic constraints.
Consistent initialization of descriptor systems naturally happens in physical
systems but should be taken into account when simulating such systems \cite{Pantelides}.
Without loss of generality, we assume that
$0<rank(\mathbf{E})=s \leq n$; $x(0)=x_{0}$ is a consistent (unknown) set of initial conditions.
If the matrix $\mathbf{E}$ is non-singular (i.e. full rank), then the descriptor form reduces to the conventional state space.
The number of algebraic constraints that must be satisfied by $x_{0}$ equals $n-s$.
We assume the pair $(\mathbf{E},A)$ to be \emph{regular}, i.e. $\det(s\mathbf{E}-A) \neq 0$ for some $s \in \mathbb{C}$
 \cite{Dai} and \emph{impulse free}, i.e. $deg \det(s\mathbf{E}-A)=rank(\mathbf{E})$ \cite{Dai}, and the triple $(\mathbf{E},A,C)$
to be observable, i.e. \cite{Ishihara}
\begin{align}
rank \left[
       \begin{array}{c}
         s\mathbf{E}-A \\
         C \\
       \end{array}
     \right]=n, \ \forall \ s \in \mathbb{C}.\notag
\end{align}
We also assume that the system (\ref{sys1})-\eqref{sys2} is locally Lipschitz with
respect to $x$ in a region $\mathcal{D}$ containing the origin,
uniformly in $u$, i.e.:
\begin{align}
&\Phi(0,u^{*})=\Psi(0,u^{*})=0,\notag\\
&\|\Phi(x_{1},u^{*})-\Phi(x_{2},u^{*})\|\leqslant\gamma_{1}\|x_{1}-x_{2}\|
,\hspace{2mm} \forall \, x_{1},x_{2}\in \mathcal{D}\notag\\
&\|\Psi(x_{1},u^{*})-\Psi(x_{2},u^{*})\|\leqslant\gamma_{2}\|x_{1}-x_{2}\|
,\hspace{2mm} \forall \, x_{1},x_{2}\in \mathcal{D}\notag
\end{align}
where $\|.\|$ is the induced 2-norm, $u^{*}$ is any admissible
control signal and $\gamma_{1}, \ \gamma_{2}>0$ are the Lipschitz constants
of $\Phi(x,u)$ and $\Psi(x,u)$, respectively, the $\mathcal{D}$ is the operating region.
If the nonlinear functions
$\Phi(x,u)$ and $\Psi(x,u)$ satisfy the Lipschitz continuity
condition globally in $\mathbb{R}^{n}$, then the results will be
valid globally. The Lipschitz continuity condition is not a restrictive assumption
since most nonlinear function are Lipschitz continuous at least locally.
Particularly, all continuously differentiable functions are known to be Lipschitz,
and their Lipschitz constant is computed as the supremum of
their Jacobian matrix over the operating region \cite{Marquez}.
As a usual assumption in robust filtering techniques, we also assume that the system under consideration
is stable.

$w(t)\in\mathcal{L}_{2}[0,\infty)$ is an unknown
exogenous disturbance, and $\Delta A(t)$ and $\Delta C(t)$ are
unknown matrices representing time-varying parameter uncertainties,
and are assumed to be of the form
\begin{eqnarray}
\left[
  \begin{array}{c}
    \Delta A(t) \\
    \Delta C(t) \\
  \end{array}
\right]= \left[
  \begin{array}{c}
    M_{1} \\
    M_{2} \\
  \end{array}
\right]F(t)N \label{uncer1}
\end{eqnarray}
where $M_{1}$, $M_{2}$ and $N$ are known real constant matrices and
$F(t)$ is an unknown real-valued time-varying matrix satisfying
\begin{equation}
F^{T}(t)F(t)\leq I \hspace{1cm} \forall \ t\in [0,\infty).
\end{equation}
The parameter uncertainty in the linear terms can be regarded as the
variation of the operating point of the nonlinear system. It is also
worth noting that the structure of parameter uncertainties in
(\ref{uncer1}) has been widely used in the problems of robust
control and robust filtering for both continuous-time and
discrete-time systems and can capture the uncertainty in a number of
practical situations \cite{deSouza1}, \cite{Khargonekar}.

\subsection{Filter Structure}
We propose the general filtering framework of the following form
\begin{equation}
\begin{split}
\left(\Sigma_{o} \right):\mathbf{E}\dot{x}_{F}(t)&=
A_{F}x_{F}(t)+B_{F}y(t)+\mathcal{E}_{1}\Phi(x_{F},u)\\
&\ \ \ +\mathcal{E}_{2}\Psi(x_{F},u)\\
z_{F}(t)&=C_{F}x_{F}(t)+D_{F}y(t)+\mathcal{E}_{3}\Psi(x_{F},u).\label{observer1}
\end{split}
\end{equation}
The proposed framework can capture both dynamic and static-gain
filter structures by proper selection of $\mathcal{E}_{1}$, $\mathcal{E}_{2}$ and
$\mathcal{E}_{3}$. Choosing $\mathcal{E}_{1}=I$, $\mathcal{E}_{2}=0$ and
$\mathcal{E}_{3}=0$ leads to the following dynamic filter
structure:
\begin{equation}
\begin{split}
\mathbf{E}\dot{x}_{F}(t)&=
A_{F}x_{F}(t)+B_{F}y(t)+\Phi(x_{F},u)\\
z_{F}(t)&=C_{F}x_{F}(t)+D_{F}y(t).\label{observer2}
\end{split}
\end{equation}
Furthermore, for the static-gain filter structure we have:
\begin{equation}
\begin{split}
\mathbf{E}\dot{x}_{F}(t)&=Ax_{F}(t)+\Phi(x_{F},u)\\
&\ \ +L[y(t)-Cx_{F}(t)-\Psi(x_{F},u)]\\
z_{F}(t)&=x_{F}(t).\label{observer3}
\end{split}
\end{equation}
Hence, with
\begin{align}
A_{F}&=A-LC, \ B_{F}=L, \ C_{F}=I, \ D_{F}=0,\notag \\
\mathcal{E}_{1}&=I, \ \mathcal{E}_{2}=-L, \ \mathcal{E}_{3}=0,\notag
\end{align}
the general filter captures the well-known static-gain observer filter
structure as a special case. We prove our result for the general
filter of class $(\Sigma_{o})$.

Now, suppose that
\begin{equation}
z(t)=Hx(t)
\end{equation}
stands for the controlled output for states to be estimated where
$H$ is a known matrix. The estimation error is defined as
\begin{multline}
e(t)\triangleq z(t)-z_{F}(t)=-C_{F}x_{F}+(H-D_{F}C-D_{F}\Delta
C)x\\-D_{F}\Psi(x,u)-\mathcal{E}_{3}\Psi(x_{F},u)-D_{F}Dw.
\label{error1}
\end{multline}

The filter error dynamics is given by
\begin{align}
\left(\Sigma_{e}
\right):\mathbf{\widetilde{E}}\dot{\xi}(t)&=(\widetilde{A}+\Delta
\widetilde{A})\xi(t)+S_{1}\Omega(\xi,u)+\widetilde{B}w(t)\\
e(t)&=(\widetilde{C}+\Delta
\widetilde{C})\xi(t)+S_{2}\Omega(\xi,u)+\widetilde{D}w(t),
\end{align}
where,
\begin{align}
&\xi\triangleq\left[
       \begin{array}{c}
         x_{F} \\
         x \\
       \end{array}
     \right], \widetilde{A}=\left[
     \begin{array}{cc}
       A_{F} & B_{F}C \\
       0 & A \\
     \end{array}
   \right], \Delta \widetilde{A}=\left[
                                   \begin{array}{cc}
                                     0 & B_{F}\Delta C \\
                                     0 & \Delta A \\
                                   \end{array}
                                 \right]\notag\\
&\mathbf{\widetilde{E}}=\left[
                          \begin{array}{cc}
                            \mathbf{E} & 0 \\
                            0 & \mathbf{E} \\
                          \end{array}
                        \right]
, \widetilde{B}=\left[
                                \begin{array}{c}
                                  B_{F}D \\
                                  B \\
                                \end{array}
                              \right], \widetilde{D}=-D_{F}D \notag\\
&\widetilde{C}=\left[
     \begin{array}{cc}
       -C_{F} & H-D_{F}C \\
     \end{array}
   \right], \Delta \widetilde{C}=\left[
                                   \begin{array}{cc}
                                     0 & -D_{F}\Delta C \\
                                   \end{array}
                                 \right]\notag\\
&\Omega(\xi,u)=\left[
                            \begin{array}{cccc}
                                \Phi(x,u) & \Psi(x,u) & \Phi(x_{F},u) & \Psi(x_{F},u)
                            \end{array}
                      \right]^{T}\notag\\
&S_{1}=\left[
         \begin{array}{cccc}
           0 & B_{F} & \mathcal{E}_{1} & \mathcal{E}_{2} \\
           I & 0     & 0 & 0 \\
         \end{array}
       \right], S_{2}=\left[
                        \begin{array}{cccc}
                          0 & -D_{F} & 0 & -\mathcal{E}_{3} \\
                        \end{array}
                      \right] \notag.
\end{align}
For the nonlinear function $\Omega$, it is easy to show that
\begin{align}
&\Gamma\triangleq\left[
                   \begin{array}{cccc}
                     0          & 0          & \gamma_{1} & \gamma_{2}\\
                     \gamma_{1} & \gamma_{2} & 0          & 0\\
                   \end{array}
                 \right]^{T}\\
&\|\Omega(\xi_{1},u)-\Omega(\xi_{2},u)\|\leq\|\Gamma(\xi_{1}-\xi_{2})\|\leq
\|\Gamma\|\|\xi_{1}-\xi_{2}\|\notag\\
&\ \
=\sqrt{\gamma_{1}^{2}+\gamma_{2}^{2}}\|\xi_{1}-\xi_{2}\|\triangleq
\gamma\|\xi_{1}-\xi_{2}\|\label{Gamma1}.
\end{align}
Thus, the filter error system is Lipschitz with Lipschitz constant
$\gamma$.

\subsection{Disturbance Attenuation Level}

Our purpose is to design the filter matrices $A_{F}$, $B_{F}$,
$C_{F}$ and $D_{F}$ such that the filter error dynamics is
asymptotically stable with maximum admissible Lipschitz constant and
the following specified $H_{\infty}$ norm upper bound is
simultaneously guaranteed.\
\begin{equation}
\|e\|\leq\mu\|w\|.
\end{equation}

In the following, we mention some useful lemmas that will be used
later in the proof of our results. \\

\emph{\textbf{Lemma 1. \cite{deSouza2}} For any
$x,y\in\mathbb{R}^{n}$ and any positive definite matrix
$P\in\mathbb{R}^{n\times{n}}$, we have}
\begin{equation}
2x^{T}y\leq x^{T}Px+y^{T}P^{-1}y.\notag
\end{equation}

\emph{\textbf{Lemma 2. \cite{deSouza2}} Let $A,D, E, F$ and P be
real matrices of appropriate dimensions with $P>0$ and $F$
satisfying $F^{T}F\leq I$. Then for any scalar $\epsilon>0$
satisfying $P^{-1}-\epsilon^{-1}DD^{T}>0$, we have}
\begin{equation}
\begin{split}
(A+DFE)^{T}P(A+DFE)\leq&
A^{T}(P^{-1}-\epsilon^{-1}DD^{T})^{-1}A+\epsilon E^{T}E.\notag
\end{split}
\end{equation}

\emph{\textbf{Lemma 3. \cite[p. 301]{Horn1}} A matrix $A \in
\mathbb{R}^{n \times n}$ is invertible if there is a matrix norm
$\||.\||$ such that $\||I-A\||<1$.}\\

The symbol $\||.\||$ in the above lemma represents any matrix norm.


\section{$H_{\infty}$ Filter Synthesis}

In this section, an $H_{\infty}$ filter with guaranteed
disturbance attenuation level $\mu$ is proposed. The
admissible Lipschitz constant is maximized through LMI optimization.
Theorem 1, introduces a design method for such a filter. It
worths mentioning that unlike the Riccati approach of
\cite{deSouza1}, in the LMI approach no $H_{\infty}$ regularity assumption is needed.\\

\emph{\textbf{Theorem 1.} Consider the Lipschitz nonlinear system
$\left(\Sigma_{s} \right)$ along with the general filter
$\left(\Sigma_{o} \right)$. The filter error dynamics is
(globally) asymptotically stable with maximum admissible
Lipschitz constant, $\gamma^{*}$, and guaranteed
$\mathfrak{L}_{2}(w \rightarrow e)$ gain, $\mu$, if there exists
a fixed scalar $\mu>0$, scalars $\epsilon_{1}>0$,
$\epsilon_{2}>0$, $\alpha_{1}>0$ and $\alpha_{2}>0$ and matrices
$C_{F}$, $D_{F}$, $P_{1}$, $P_{2}$, $G_{1}$, $G_{2}$ and $\mathcal{E}_{3}$ such that
the following optimization problem has a solution.}
\begin{align}
&\hspace{.5cm} \min (2\alpha_{1}+\alpha_{2}) \notag\\
&\text{s.t.}\notag\\
&\Xi_{1}=\left[
           \begin{array}{ccccc}
             \Pi_{1} & \Pi_{2} & 0        & \Pi_{3} & \Pi_{4} \\
             \star   & \Pi_{5} & \Pi_{6}  & 0       & 0 \\
             \star   & \star   & \Pi_{7}  & 0       & 0 \\
             \star   & \star   & \star    & \Pi_{8} & 0 \\
             \star   & \star   & \star    & \star   & \Pi_{9} \\
           \end{array}
         \right]<0\label{LMI1}\\
&\Xi_{2}=\left[
   \begin{array}{ccc}
     \epsilon_{2}I & 0 & -D_{F}M_{2} \\
     \star & I & 0 \\
     \star & \star & I \\
   \end{array}
 \right]>0\label{LMI2}\\
&\Xi_{3}=\left[
   \begin{array}{ccc}
     \alpha_{1}I & \mathcal{E}_{3} & D_{F}\\
     \star & \alpha_{1}I & 0 \\
     \star & \star & \alpha_{1}I
   \end{array}
 \right]>0\label{LMI3}\\
&\Xi_{4}=\left[
           \begin{array}{cc}
             I & I-P_{1}^{T} \\
             \star & I \\
           \end{array}
         \right]>0\label{non-sin}\\
&\mathbf{E}^{T}P_{1}=P_{1}^{T}\mathbf{E}\geq 0 \label{E1}\\
&\mathbf{E}^{T}P_{2}=P_{2}^{T}\mathbf{E}\geq 0 \label{E2}
\end{align}
\emph{where the elements of $\Xi_{1}$ are as defined in the following,
$\Lambda_{1}=G_{1}^{T}+G_{1}$,
$\Lambda_{2}=A^{T}P_{2}+P_{2}A+(\epsilon_{1}+\epsilon_{2})N^{T}N$
and $\Lambda_{3}=H^{T}-C^{T}D_{F}^{T}$.
\begin{align}
&\Pi_{1}=\left[
           \begin{array}{ccc}
             \Lambda_{1} & G_{2}C & I \\
             \star & \Lambda_{2} & 0 \\
             \star & \star & -\alpha_{2}I \\
           \end{array}
         \right], \Pi_{4}=\left[
                                             \begin{array}{cc}
                                                G_{2}D & 0 \\
                                                P_{2}B & 0 \\
                                                0 & 0 \\
                                             \end{array}
                                            \right],\notag\\
&\Pi_{2}=\left[
                            \begin{array}{ccc}
                              0 & G_{2}M_{2} & -C_{F}^{T} \\
                              0 & P_{2}M_{1} & \Lambda_{3} \\
                              0 & 0 & 0 \\
                            \end{array}
                          \right], \Pi_{6}=\left[
           \begin{array}{cc}
             0 & 0 \\
             0 & 0 \\
             0 & -D_{F}M_{2} \\
           \end{array}
         \right],\notag\\
&\Pi_{3}=\left[
           \begin{array}{cccc}
             0 & G_{2} & P_{1}\mathcal{E}_{1} & P_{1}\mathcal{E}_{2} \\
             P_{2} &0 & 0 & 0 \\
             0& 0 & 0 & 0 \\
           \end{array}
         \right],\notag\\
&\Pi_{5}=diag(-\epsilon_{1}I, -\epsilon_{1}I,-\frac{1}{3}I), \Pi_{7}=diag(-\frac{1}{3}\epsilon_{2}I,-\frac{1}{3}\epsilon_{2}I), \notag\\ &\Pi_{8}=diag(-I,-I,-I,-I), \Pi_{9}=\left[
           \begin{array}{cc}
             -\mu^{2}I & -D^{T}D_{F}^{T}\\
             \star & -\frac{1}{3}I \\
           \end{array}
         \right]\notag.
\end{align}
Once the problem is solved:}
\begin{align}
A_{F}&=P^{-1}_{1}G_{1},\label{AF}\\
B_{F}&=P^{-1}_{1}G_{2},\label{BF}\\
C_{F}\ &\text{and}\ D_{F}\ \text{are directly obtained},\notag
\\\alpha_{1}^{*} &\triangleq \min(\alpha_{1}),
\\\alpha_{2}^{*} &\triangleq \min(\alpha_{2}),
\\\gamma^{*} &\triangleq \max(\gamma)=\frac{1}{\sqrt{\alpha_{2}^{*}(1+3{\alpha_{1}^{*}}^{2})}}.
\end{align}\\
\textbf{Proof:} Consider the following Lyapunov function candidate
\begin{align}
V(\xi(t))=\xi^{T}\mathbf{\widetilde{E}}^{T}P\xi.
\end{align}
To prove the stability of the filter error dynamics, we employ the
well-established generalized Lyapunov stability theory as discussed
in \cite{He_Wang}, \cite{Masubuchi} and \cite{Ishihara} and the
references therein. The generalized Lyapunov stability theory is
mainly based on an extended version of the well-known LaSalle's
invariance principle for descriptor systems. Based on this theory,
the above function along with the conditions \eqref{E1} and
\eqref{E2} is a generalized Lyapunov function (GLF) for the system
$\left(\Sigma_{e} \right)$ where $P=diag(P_{1},P_{2})$. In fact, it
can be shown that $V(\xi(t))=0$ if and only if
$\mathbf{\widetilde{E}}\xi=0$  and positive
elsewhere \cite[Ch. 2]{He_Wang}. Now, we calculate the derivative of $V$ along the
trajectories of $\left(\Sigma_{e} \right)$. We have
\begin{equation}
\begin{split}
\dot{V}&=\dot{\xi}^{T}\mathbf{\widetilde{E}}^{T} P\xi+\xi^{T}
\mathbf{\widetilde{E}}^{T}P\dot{\xi}=2\xi^{T}(\widetilde{A}+\Delta
\widetilde{A})^{T}P\xi\\
&\ \ \ +2\xi^{T}PS_{1}\Omega+2\xi^{T}P\widetilde{B}w.\label{Vdot}
\end{split}
\end{equation}
Now, we define
\begin{equation}
J\triangleq \int^{\infty}_{0}(e^{T}e-\mu^{2} w^{T}w) dt.
\end{equation}
Therefore
\begin{equation}
J<\int^{\infty}_{0}(e^{T}e-\mu^{2} w^{T}w+\dot{V}) dt
\end{equation}
so a sufficient condition for $J\leq0$ is that
\begin{equation}
\forall t\in[0,\infty),\hspace{5mm} e^{T}e-\mu^{2}
w^{T}w+\dot{V}\leq0\label{J1}.
\end{equation}
We have
\begin{equation}
\begin{split}
e^{T}e&=\xi^{T}(\widetilde{C}+\Delta
\widetilde{C})^{T}(\widetilde{C}+\Delta
\widetilde{C})\xi+2\xi^{T}(\widetilde{C}+\Delta
\widetilde{C})^{T}S_{2}\Omega\\
&\ \ +2\xi^{T}(\widetilde{C}+\Delta\widetilde{C})^{T}\widetilde{D}w\notag+2w^{T}\widetilde{D}^{T}S_{2}\Omega\\
&\ \
+\Omega^{T}S_{2}^{T}S_{2}\Omega+w^{T}\widetilde{D}^{T}\widetilde{D}w
\end{split}
\end{equation}
Thus, using Lemma 1,
\begin{equation}
\begin{split}
\dot{V}&+e^{T}e-\mu^{2} w^{T}w\leq2\xi^{T}(\widetilde{A}+\Delta
\widetilde{A})^{T}P\xi+2\xi^{T}PS_{1}\Omega\\
&+2\xi^{T}P\widetilde{B}w-\mu^{2}w^{T}w+2\xi^{T}(\widetilde{C}+\Delta
\widetilde{C})^{T}(\widetilde{C}+\Delta \widetilde{C})\xi \\
&+2\xi^{T}(\widetilde{C}+\Delta\widetilde{C})^{T}\widetilde{D}w+3\Omega^{T}S_{2}^{T}S_{2}\Omega+2w^{T}\widetilde{D}^{T}\widetilde{D}w\\
&\leq\xi^{T}[2(\widetilde{A}+\Delta
\widetilde{A})^{T}P+PS_{1}^{T}S_{1}P+2(\widetilde{C}+\Delta
\widetilde{C})^{T}\cdots\notag
\end{split}
\end{equation}
\begin{equation}
\begin{split}
&\cdots(\widetilde{C}+\Delta
\widetilde{C})]\xi+\xi^{T}[2P\widetilde{B}+2(\widetilde{C}+\Delta
\widetilde{C})^{T}\widetilde{D}]w\\
&+\Omega^{T}\Omega+3\Omega^{T}S_{2}^{T}S_{2}\Omega+w^{T}(2\widetilde{D}^{T}\widetilde{D}-\mu^{2}I)w\\
&\leq\xi^{T}[2(\widetilde{A}+\Delta
\widetilde{A})^{T}P+PS_{1}^{T}S_{1}P+3(\widetilde{C}+\Delta
\widetilde{C})^{T}\cdots\\
&\cdots(\widetilde{C}+\Delta
\widetilde{C})]\xi+2\xi^{T}P\widetilde{B}w+\Omega^{T}\Omega+3\Omega^{T}S_{2}^{T}S_{2}\Omega\\
&+w^{T}(3\widetilde{D}^{T}\widetilde{D}-\mu^{2}I)w\label{Vdot4}.
\end{split}
\end{equation}
Without loss of generality, we assume that there is a scalar $\alpha_{1}$ such that $\|\mathcal{E}_{3}\mathcal{E}_{3}^{T}+D_{F}D_{F}^{T}\|< \alpha_{1}^{2}$
where, $\alpha_{1}>0$ is an unknown variable. Thus,
\begin{equation}
\begin{split}
\Omega^{T}\Omega&+3\Omega^{T}S_{2}^{T}S_{2}\Omega \leq
(1+3\|S_{2}^{T}S_{2}\|)\Omega^{T}\Omega\\
&=(1+3\|S_{2}S_{2}^{T}\|)\Omega^{T}\Omega\\
&=[1+3\|\mathcal{E}_{3}\mathcal{E}_{3}^{T}+D_{F}D_{F}^{T}\|]\Omega^{T}\Omega\\
&<(1+3\alpha_{1}^{2})\Omega^{T}\Omega \\
&\leq(1+3\alpha_{1}^{2})\xi^{T}\Gamma^{T}\Gamma\xi\\
&\leq(1+3\alpha_{1}^{2})\gamma^{2}\xi^{T}\xi.\label{Vdot1}
\end{split}
\end{equation}
Note that $\Omega(0,u)=0$. Now, defining the change of variables
\begin{align}
\alpha_{2}&\triangleq
\frac{1}{(1+3\alpha_{1}^{2})\gamma^{2}} \Rightarrow\gamma=\frac{1}{\sqrt{\alpha_{2}(1+3{\alpha_{1}}^{2})}}\label{Vdot2},
\end{align}
we have
\begin{align}
\Omega^{T}\Omega+3\Omega^{T}S_{2}^{T}S_{2}\Omega <
\alpha_{2}^{-1}\xi^{T}\xi.
\end{align}
It is worth mentioning that the change of variables in \eqref{Vdot2}
plays a vital role here. The alternative changes of variables such
as $\alpha_{2}=\alpha_{1} \gamma$ which may seem more
straightforward, would make $\gamma$ appear in $\Xi_{1}$ and then
due to the existence of $\alpha_{1}$ in the LMI \eqref{LMI3}, the
variables $\alpha_{1}$ and $\gamma$ would be over-determined.\\
On the other hand,
\begin{align}
\Delta \widetilde{A}&=\left[
                                   \begin{array}{cc}
                                     0 & B_{F}\Delta C \\
                                     0 & \Delta A \\
                                   \end{array}
                                 \right]=\left[
                                           \begin{array}{cc}
                                             0 & B_{F}M_{2}FN \\
                                             0 & M_{1}FN \\
                                           \end{array}
                                         \right]\notag\\&=\left[
                                           \begin{array}{cc}
                                             0 & B_{F}M_{2}\\
                                             0 & M_{1}\\
                                           \end{array}
                                         \right]F\left[
                                                   \begin{array}{cc}
                                                     0 & 0 \\
                                                     0 & N \\
                                                   \end{array}
                                                 \right]\triangleq \widetilde{M_{1}}F\widetilde{N}\label{DAT}\\
\Delta \widetilde{C}&=\left[
                                   \begin{array}{cc}
                                     0 & -D_{F}\Delta C \\
                                   \end{array}
                                 \right]=\left[
                                   \begin{array}{cc}
                                     0 & -D_{F}M_{2}FN \\
                                   \end{array}
                                 \right]\notag\\&=\left[
                                   \begin{array}{cc}
                                     0 & -D_{F}M_{2} \\
                                   \end{array}
                                 \right]F\left[
                                                   \begin{array}{cc}
                                                     0 & 0 \\
                                                     0 & N \\
                                                   \end{array}
                                                 \right]\triangleq
                                                 \widetilde{M_{2}}F\widetilde{N}.\label{DCT}
\end{align}
Therefore, based on \eqref{Vdot1} and \eqref{Vdot2} and using Lemma
2 we can write
\begin{equation}
\begin{split}
\dot{V}&+e^{T}e-\mu^{2} w^{T}w<
\xi^{T}[\widetilde{A}^{T}P+P\widetilde{A}+\epsilon_{1}\widetilde{N}^{T}N\\
&+\epsilon_{1}^{-1}P\widetilde{M}_{1}\widetilde{M}_{1}P+3\widetilde{C}^{T}(I-\epsilon_{2}^{-1}\widetilde{M}_{2}\widetilde{M}_{2}^{T})^{-1}\widetilde{C}\\
&+\epsilon_{2}\widetilde{N}^{T}N+PS_{1}S_{1}^{T}P+\alpha_{2}^{-1}]\xi+2\xi^{T}P\widetilde{B}w\\
&+w^{T}(3\widetilde{D}^{T}\widetilde{D}-\mu^{2}I)w.\label{Vdot3}
\end{split}
\end{equation}
Now, a sufficient condition for \eqref{J1} is that the right hand
side of \eqref{Vdot3} be negative definite. Using Schur complements,
this is equivalent to the following LMI. Note that having $w=0$,
\eqref{Vdot} is already included in \eqref{Vdot4} and consequently
in \eqref{Vdot3}.

{\footnotesize \begin{align} &\left[
  \begin{array}{cccccccc}
    \Upsilon & I & P\widetilde{M}_{1} & \widetilde{C}^{T} & 0 & PS_{1} & P\widetilde{B} & 0 \\
    \star & -\alpha_{2}I & 0 & 0 & 0 & 0 & 0 & 0 \\
    \star & \star & -\epsilon_{1}I & 0 & 0 & 0 & 0 & 0 \\
    \star & \star & \star & -\frac{1}{3}I & \widetilde{M}_{2} & 0 & 0 & 0 \\
    \star & \star & \star & \star & -\frac{\epsilon_{2}}{3}I & 0 & 0 & 0 \\
    \star & \star & \star & \star & \star & -I & 0 & 0 \\
    \star & \star & \star & \star & \star & \star & -\mu^{2}I & \widetilde{D}^{T} \\
    \star & \star & \star & \star & \star & \star & \star & -\frac{1}{3}I \\
  \end{array}
\right]<0\notag\\
&\Upsilon=\widetilde{A}^{T}P+P\widetilde{A}+(\epsilon_{1}+\epsilon_{2})\widetilde{N}^{T}\widetilde{N}\notag
\end{align}}
Substituting from \eqref{DAT} and \eqref{DCT}, having
$P=diag(P_{1},P_{2})$, defining change of variables $G_{1}\triangleq
P_{1}A_{F}$ and $G_{2}\triangleq P_{1}B_{F}$ and using Schur
complements, the LMI \eqref{LMI1} is obtained. The LMI \eqref{LMI2}
is equivalent to the condition
$I-\epsilon_{2}^{-1}\widetilde{M}_{2}\widetilde{M}_{2}^{T}>0$ needed
in Lemma 2. Now we return to the condition $\|E_{2}E_{2}^{T}+D_{F}D_{F}^{T}\|< \alpha_{1}^{2}$; we have
\begin{align}
&\|\mathcal{E}_{3}\mathcal{E}_{3}^{T}+D_{F}D_{F}^{T}\|< \alpha_{1}^{2} \Rightarrow \left\|\left[
                                                                  \begin{array}{cc}
                                                                    \mathcal{E}_{3} & D_{F} \\
                                                                  \end{array}
                                                                \right]\left[
\begin{array}{c}
    \mathcal{E}_{3}^{T} \\
    D_{F}^{T} \\
  \end{array}
\right]
 \right\|< \alpha_{1}^{2} \notag \\
 &\Rightarrow \left\|\left[
                                                                \begin{array}{cc}
                                                                    \mathcal{E}_{3} & D_{F} \\
                                                                  \end{array}
                                                                \right]\right\| < \alpha_{1},
\end{align}

which by means of Schur complement lemma is equivalent to the LMI \eqref{LMI3}.\\
Note that neither $P_{1}$ nor $P_{2}$ are necessarily positive
definite. However, in order to Find $A_{F}$ and $B_{F}$ in
\eqref{AF} and \eqref{BF}, $P_{1}$ must be invertible. Since we are
using the spectral matrix norm (matrix 2-norm) through out this
work, based on Lemma 3, a sufficient condition for nonsingularity
of $P_{1}$ is that $\|I-P_{1}\|=\sigma_{max}(I-P_{1})<1$. This is
equivalent to $I-(I-P_{1})^{T}(I-P_{1})>0$. Thus, using Schur's
complement, LMI \eqref{non-sin} guarantees the nonsingularity of $P_{1}$.\\
Now, maximization of $\gamma$ can be done by the simultaneous
minimization of $\alpha_{1}$ and $\alpha_{2}$. In order to cast it
in the form of an LMI optimization problem, combining the two
objective functions we minimize the scalarized linear objective
function $c_{1}\alpha_{1}+c_{2}\alpha_{2}$. To determine the weights
$c_{1}$ and $c_{2}$ in the objective function, we compute the
sensitivity of $\gamma$ to the changes of $\alpha_{1}$ and
$\alpha_{2}$. We have
\begin{align}
\mathcal{S}^{\gamma}_{\alpha_{1}}&=\frac{\partial
\gamma}{\partial\alpha_{1}}.\frac{\alpha_{1}}{\gamma}=\frac{-3\alpha_{1}^{2}}{1+3\alpha_{1}^{2}}>-1\\
\mathcal{S}^{\gamma}_{\alpha_{2}}&=\frac{\partial
\gamma}{\partial\alpha_{2}}.\frac{\alpha_{2}}{\gamma}=-\frac{1}{2}.
\end{align}
Hence, $\gamma$ is up to twice more sensitive to the changes of
$\alpha_{1}$ than those of $\alpha_{2}$. Note that the absolute
value of the sensitivity function determines the amount of
sensitivity while its sign determines the direction of the
sensitivity. So, a reasonable choice can be $c_{1}=2$ and $c_{2}=1$. $\blacksquare$\\

\textbf{Remark 1.} Maximization of $\gamma$ guarantees the robust
asymptotic stability of the error filter dynamics for any
Lipschitz nonlinear function $\Omega(\xi,u)$ with Lipschitz constant
less than or equal $\gamma^{*}$. It is clear that if a filter for
a system with a given fixed Lipschitz constant is to be designed,
the proposed LMI optimization problem will reduce to an LMI
feasibility problem and there is no need for the change of
variable \eqref{Vdot2} anymore.\\

\textbf{Remark 2.} The proposed LMIs are linear in $\alpha_{1}$,
$\alpha_{2}$ and $\zeta(=\mu^{2})$. Thus, either can be a fixed
constant or an optimization variable. So, either the admissible
Lipschitz constant or the disturbance attenuation level can be
considered as an optimization variable in Theorem 1. Given this, it
may be more realistic to have a combined performance index. This
leads to a multiobjective convex optimization problem optimizing
both $\gamma$ and $\mu$, simultaneously. See \cite{Abbaszadeh3} and
\cite{Abbaszadeh5} for details and examples of multiobjective
optimization approach to $H_{\infty}$ filtering for other classes
of nonlinear systems.\\

Note that $\mathcal{E}_{1}$ and $\mathcal{E}_{2}$ are not
optimization variables. They are \emph{apriory} fixed constant
matrices that determine the structure of the filter while
$\mathcal{E}_{3}$ can be either a fixed gain or an optimization variable.
It is worth mentioning that in the case of static-gain filter, some
simplification can be made. First of all, since in this structure
$D_{F}=0$, the LMIs \eqref{LMI2} and \eqref{LMI3} are eliminated.
Besides, since $\mathcal{E}_{3}=0$, the inequality \eqref{Vdot1}
reduces to $\Omega^{T}\Omega+2\Omega^{T}S_{2}^{T}S_{2}\Omega\leq
\gamma^{2}\xi^{T}\xi$ and there is no need to the change of
variables \eqref{Vdot2}. Consequently, the cost function simplifies
to $\max(\gamma)$. In addition, for this structure we have
\begin{align}
A_{F}&=A-LC \Rightarrow P_{1}A_{F}=P_{1}A-P_{1}LC\\
B_{F}&=L, \mathcal{E}_{2}=-L \Rightarrow
P_{1}B_{F}=-P_{1}\mathcal{E}_{2}=P_{1}L.
\end{align}
Therefore, instead of variables $G_{1}$ and $G_{2}$, a change of
variables $G=P_{1}L$ is enough. Obviously, the dynamic filter
structure has more degrees freedom and can provide a robust filter
in some of the cases for which a static-gain filter does not
exist.

\section{Converting SDP into strict LMIs}
Due to the existence of equalities and non-strict inequalities in
\eqref{E1} and \eqref{E2}, the optimization problem of Theorem 1 is
not a convex \emph{strict} LMI Optimization and instead it is a
Semidefinite Programming (SDP) with quasi-convex solution space. The
SDP problem proposed in Theorem 1 can be solved using freely
available packages such as YALMIP \cite{YALMIP} or SeDuMi \cite{SeDuMi}. However, in order to use the
numerically more efficient Matlab strict LMI solver, in this section we convert the SDP
problem proposed in Theorem 1 into a strict LMI optimization
problem through a smart transformation. We use a similar approach as
used in \cite{Uezato} and \cite{Lu_G3}. Let $\mathbf{E}_{\bot} \in
\mathbb{R}^{(n-s)\times n}$ be the orthogonal complement of
$\mathbf{E}$ such that $\mathbf{E}_{\bot}\mathbf{E}=0$ and
$\text{rank}(\mathbf{E}_{\bot})=n-s$. The following corollary gives
the strict LMI formulation.\\

\emph{\textbf{Corollary 1.} Consider the Lipschitz nonlinear system
$\left(\Sigma_{s} \right)$ along with the general filter
$\left(\Sigma_{o} \right)$. The filter error dynamics is
(globally) asymptotically stable with maximum admissible
Lipschitz constant, $\gamma^{*}$, and guaranteed
$\mathfrak{L}_{2}(w \rightarrow e)$ gain, $\mu$, if there exists a
$\mu>0$, scalars $\epsilon_{1}>0$,
$\epsilon_{2}>0$, $\alpha_{1}>0$ and $\alpha_{2}>0$ and matrices
$C_{F}$, $D_{F}$, $X_{1}>0$, $X_{2}>0$, $Y_{1}$, $Y_{2}$, $G_{1}$,
$G_{2}$ and $\mathcal{E}_{3}$ such that the following LMI optimization problem has a
solution.}\\
\begin{align}
&\hspace{.3cm} \min (2\alpha_{1}+\alpha_{2}) \notag\\
&\hspace{-.3cm}\text{s.t.}\notag\\
&\Xi_{1}<0 \notag\\
&\Xi_{2}>0\notag\\
&\Xi_{3}>0\notag\\
&\Xi_{4}=\left[
           \begin{array}{cc}
             I & I-P_{1}^{T} \\
             \star & I \\
           \end{array}
         \right]>0,\notag
\end{align}
\emph{where, $\Xi_{1}$, $\Xi_{2}$, $\Xi_{3}$ and $\Xi_{4}$ are as in
Theorem 1 with}
\begin{align}
P_{1}&=X_{1}\mathbf{E}+\mathbf{E}_{\bot}^{T}Y_{1}\label{P1}\\
P_{2}&=X_{2}\mathbf{E}+\mathbf{E}_{\bot}^{T}Y_{2}\label{P2}.
\end{align}
\emph{Once the problem is solved:}
\begin{align}
&A_{F}=P^{-1}_{1}G_{1}=(X_{1}\mathbf{E}+\mathbf{E}_{\bot}^{T}Y_{1})^{-1}G_{1},\label{AF2}\\
&B_{F}=P^{-1}_{1}G_{2}=(X_{1}\mathbf{E}+\mathbf{E}_{\bot}^{T}Y_{1})^{-1}G_{2},\label{BF2}\\
&C_{F}\ \text{and}\ D_{F}\ \text{are directly obtained},\notag
\\&\alpha_{1}^{*} \triangleq \min(\alpha_{1}), \ \ \alpha_{2}^{*} \triangleq \min(\alpha_{2}),\notag
\\&\gamma^{*} \triangleq \max(\gamma)=\frac{1}{\sqrt{\alpha_{2}^{*}(1+3{\alpha_{1}^{*}}^{2})}}.\notag
\end{align}

\textbf{Proof:} We have
$\mathbf{E}^{T}P_{1}=\mathbf{E}^{T}(X_{1}\mathbf{E}+\mathbf{E}^{T}_{\perp}Y)=\mathbf{E}^{T}X_{1}\mathbf{E}$.
Since $X_{1}$ is positive definite, $\mathbf{E}^{T}X_{1}\mathbf{E}$
is always at least positive semidefinite (and thus symmetric), i.e.
$\mathbf{E}^{T}P_{1}=P_{1}^{T}\mathbf{E}\geq 0$. Similarly, we have
$\mathbf{E}^{T}P_{2}=P_{2}^{T}\mathbf{E}=\mathbf{E}^{T}X_{2}\mathbf{E}\geq 0$.
Therefore, the two conditions \eqref{E1} and \eqref{E2} are
included in \eqref{P1} and \eqref{P2}.
Now suppose $\widetilde{X}=diag (X_{1},X_{2})$ and
$P=diag(P_{1},P_{2})$. We have
\begin{align}
V=\xi^{T} \mathbf{\widetilde{E}}^{T}P
\xi=\xi^{T}\mathbf{\widetilde{E}}^{T}\widetilde{X}\mathbf{\widetilde{E}}\xi.
\end{align}
Since $X_{1}$ and $X_{2}$ are positive definite, so is
$\widetilde{X}$. Hence, $V$ is always greater than zero and vanishes
if and only if $\mathbf{\widetilde{E}}\xi=0$. Thus, the
transformations \eqref{P1} and \eqref{P2} preserve the legitimacy of
$V$ as a generalized Lyapunov function for the filter error
dynamics. The rest of the proof is the same as the proof of Theorem
1. $\blacksquare$\\

\textbf{Remark 3.} The beauty of the above result is that with a smart
change of variables, the quasi-convex semidefinite programming problem
is converted into a convex strict LMI optimization without any
approximation. Although theoretically the two problems are
equivalent, numerically, the strict LMI optimization problem can be
solved more efficiently. Note that by replacing $P_{1}$ and $P_{2}$
from \eqref{P1} and \eqref{P2} into $\Xi_{1}$ and solving the LMI
optimization problem of Corollary 1, the matrices $X_{1}$, $X_{2}$,
$Y_{1}$ and $Y_{2}$ are directly obtained. Then, having the
nonsingularity of $P_{1}$ guaranteed, the two matrices $A_{F}$ and
$B_{F}$ are obtained as given in \eqref{AF2} and \eqref{BF2},
respectively.\\

The LMI optimization problems are convex in the decision variables
and the strict LMI solvers are well known to be very
efficent and of low computational complexity \cite{Boyd}.
The interior-point methods exploited to solve LMIs
are scalable to large problems. Although the complexity of LMI computations
can grow quickly with the problem order (number of states),
but it is still much lower than equivalent SDP computations.
We also emphasize that the design procedure proposed in this work is \emph{offline},
and thus, the computational burden is not restricted to real-time implementation aspects such
as sampling time.

In the next section we discuss an important feature of the proposed
filter, robustness against nonlinear uncertainty.

\section{Robustness Against Nonlinear Uncertainty}
As mentioned earlier, the maximization of Lipschitz constant makes
the proposed filter robust against some Lipschitz nonlinear
uncertainty. In this section this robustness feature is studied and
norm-wise bounds on the nonlinear uncertainty are derived. The
norm-wise analysis provides an upper bound on the Lipschitz constant
of the nonlinear uncertainty and the norm of the Jacobian matrix of
the corresponding nonlinear function.

Assume a nonlinear uncertainty as follows
\begin{align}
&\Phi_{\Delta}(x,u)=\Phi(x,u)+\Delta\Phi(x,u)\label{uncer2}\\
&\Psi_{\Delta}(x,u)=\Psi(x,u)+\Delta\Psi(x,u)\label{uncer3}\\
&\dot{x}(t)=(A+ \Delta A)x(t)+\Phi_{\Delta}(x,u)+Bw(t)\\
&y(t)=(C+ \Delta C)x(t)+\Psi_{\Delta}(x,u)+Dw(t),
\end{align}
where $\Phi_{\Delta}$ and $\Psi_{\Delta}$ are uncertain nonlinear
functions and $\Delta\Phi$ and $\Delta\Psi$ are unknown nonlinear
uncertainties. Suppose that
\begin{align}
&\|\Delta\Phi(x_{1},u)-\Delta\Phi(x_{2},u)\|\leqslant\Delta\gamma_{1}\|x_{1}-x_{2}\|, \ \forall x_{1},x_{2} \in \mathcal{D},\notag\\
&\|\Delta\Psi(x_{1},u)-\Delta\Psi(x_{2},u)\|\leqslant\Delta\gamma_{2}\|x_{1}-x_{2}\|,\ \forall x_{1},x_{2} \in \mathcal{D}.\notag
\end{align}

Estimating and modeling nonlinear uncertainty can be made through nonlinear system
identification techniques as well as numerical Monte-Carlo simulations \cite{Luck.etal2004,McKenna.etal2004}.
For physical (first-principle) models, bounds on the uncertain are often associated with the physical knowledge
about the range of variations in the model parameters (see for example \cite{Rehman.etal2011,Agamennonia.etal2004}).
Therefore, for descriptor systems derived via physical modeling, if the nominal part of the system is Lipschitz,
which is often the case, it is reasonable to assume the nonlinearity as being Lipschitz, as well.
For emprical/statistical models, these bounds are estimated using the experimental data and
rigorous simulations of possible scenarios \cite{Luck.etal2004,McKenna.etal2004}.
Certain properties of nonlinear uncertainties (such as Lipschitz continuity)
can also be verified based the domain-based physical knowledge or statistical methods \cite{McKenna.etal2004,Kosut1997}.

\emph{\textbf{Proposition 1.}} {\emph{Suppose that the actual
Lipschitz constant of the nonlinear functions $\Phi$ and $\Psi$ are
$\gamma_{1}$ and $\gamma_{2}$, respectively and the maximum
admissible Lipschitz constant achieved by Theorem 1 (Corollary 1), is
$\gamma^{*}$. Then, the filter designed based on Theorem 1 (Corollary 1), can
tolerate any additive Lipschitz nonlinear uncertainties over $\Phi$
and $\Psi$ with Lipschitz constants $\Delta \gamma_{1}$ and $\Delta
\gamma_{2}$ such that $\sqrt{\left(\gamma_{1}+\Delta
\gamma_{1}\right)^{2}+\left(\gamma_{2}+\Delta
\gamma_{2}\right)^{2}}\leq\gamma^{*}$}}.\\
\\
\textbf{Proof:} We have,
\begin{align}
\Omega_\Delta(\xi,u)&=\Omega(\xi,u)+\Delta\Omega(\xi,u)=\left[
                                                         \begin{array}{c}
                                                           \Phi_{\Delta}(x,u)\\
                                                           \Psi_{\Delta}(x,u) \\
                                                           \Phi_{\Delta}(x_{F},u) \\
                                                           \Psi_\Delta(x_{F},u) \\
                                                         \end{array}
                                                       \right]=\left[
                                                         \begin{array}{c}
                                                           \Phi(x_{F},u)\\
                                                           \Psi(x,u) \\
                                                           \Phi(x_{F},u) \\
                                                           \Psi(x_{F},u) \\
                                                         \end{array}
                                                       \right]+\left[
                                                         \begin{array}{c}
                                                           \Delta\Phi(x,u) \\
                                                           \Delta\Psi(x,u) \\
                                                           \Delta\Phi(x_{F},u) \\
                                                           \Delta\Psi(x_{F},u) \\
                                                         \end{array}
                                                       \right].\notag
\end{align}
Based on Schwarz inequality,
\begin{eqnarray}
\begin{split}
\|\Phi_{\Delta}(x_{1},u)&-\Phi_{\Delta}(x_{2},u)\|\leq
\|\Phi(x_{1},u)-\Phi(x_{2},u)\|\\
&+\|\Delta\Phi(x_{1},u)-\Delta\Phi(x_{2},u)\|\leq \gamma_{1}\|x_{1}-x_{2}\|\\
&+\Delta\gamma_{1}\|x_{1}-x_{2}\|=(\gamma_{1}+\Delta\gamma_{1})\|x_{1}-x_{2}\|.
\end{split}\notag
\end{eqnarray}
Similarly,
\begin{align}
\|\Psi_{\Delta}(x_{1},u)-\Psi_{\Delta}(x_{2},u)\|\leq
(\gamma_{2}+\Delta\gamma_{2})\|x_{1}-x_{2}\|.
\end{align}
Based on \eqref{Gamma1}, we can write
\begin{align}
&\Gamma_{\Delta}\triangleq\left[
                   \begin{array}{cc}
                     0 & \gamma_{1}+\Delta \gamma_{1}\\
                     0 & \gamma_{2}+\Delta \gamma_{2}\\
                     \gamma_{1}+\Delta \gamma_{1} & 0 \\
                     \gamma_{2}+\Delta \gamma_{2} & 0 \\
                   \end{array}
                 \right]\label{Gamma_Delta1}\\
&\|\Omega_\Delta(\xi_{1},u)-\Omega_\Delta(\xi_{2},u)\|\leq\|\Gamma_\Delta(\xi_{1}-\xi_{2})\|\notag\\
&\hspace{1cm}\leq \sqrt{\left(\gamma_{1}+\Delta
\gamma_{1}\right)^{2}+\left(\gamma_{2}+\Delta
\gamma_{2}\right)^{2}}\|\xi_{1}-\xi_{2}\|.
\end{align}
On the other hand, according to the Theorem 1,
$\Omega_{\Delta}(x,u)$ can be any Lipschitz nonlinear function with
Lipschitz constant less than or equal to $\gamma^{*}$,
\begin{equation}
\|\Omega_{\Delta}(\xi_{1},u)-\Omega_{\Delta}(\xi_{2},u)\|\leq\gamma^{*}\|\xi_{1}-\xi_{2}\|,\notag
\end{equation}
so, there must be
\begin{eqnarray}
\sqrt{\left(\gamma_{1}+\Delta
\gamma_{1}\right)^{2}+\left(\gamma_{2}+\Delta
\gamma_{2}\right)^{2}}\leq{\gamma^{*}}\label{bound1}.\;\: \ \
\blacksquare
\end{eqnarray}
In addition, we know that if $\Delta\Phi$ and $\Delta\Psi$ are
continuously differentiable functions on $\mathcal{D}$, then
$\forall x, x_{1}, x_{2} \in \mathcal{D}$,
\begin{align}
\|\Delta\Phi(x_{1},u)-\Delta\Phi(x_{2},u)\|&\leqslant\|\frac{\partial\Delta\Phi}{\partial
x}(x_{1}-x_{2})\|,\notag\\
\|\Delta\Psi(x_{1},u)-\Delta\Psi(x_{2},u)\|&\leqslant\|\frac{\partial\Delta\Psi}{\partial
x}(x_{1}-x_{2})\|,\notag
\end{align}
where $\frac{\partial\Delta\Phi}{\partial x}$ and
$\frac{\partial\Delta\Psi}{\partial x}$ are the Jacobian matrices
\cite{Marquez}. So $\Delta\Phi(x,u)$ and $\Delta\Psi(x,u)$ can be
any additive uncertainties with
$\sqrt{(\gamma_{1}+\|\frac{\partial\Delta\Phi}{\partial
x}\|)^{2}+(\gamma_{2}+\|\frac{\partial\Delta\Psi}{\partial
x}\|)^{2}}\leq{\gamma^{*}}$.\\

\begin{figure}[!h]
  \centering
  \includegraphics[width=4.5in]{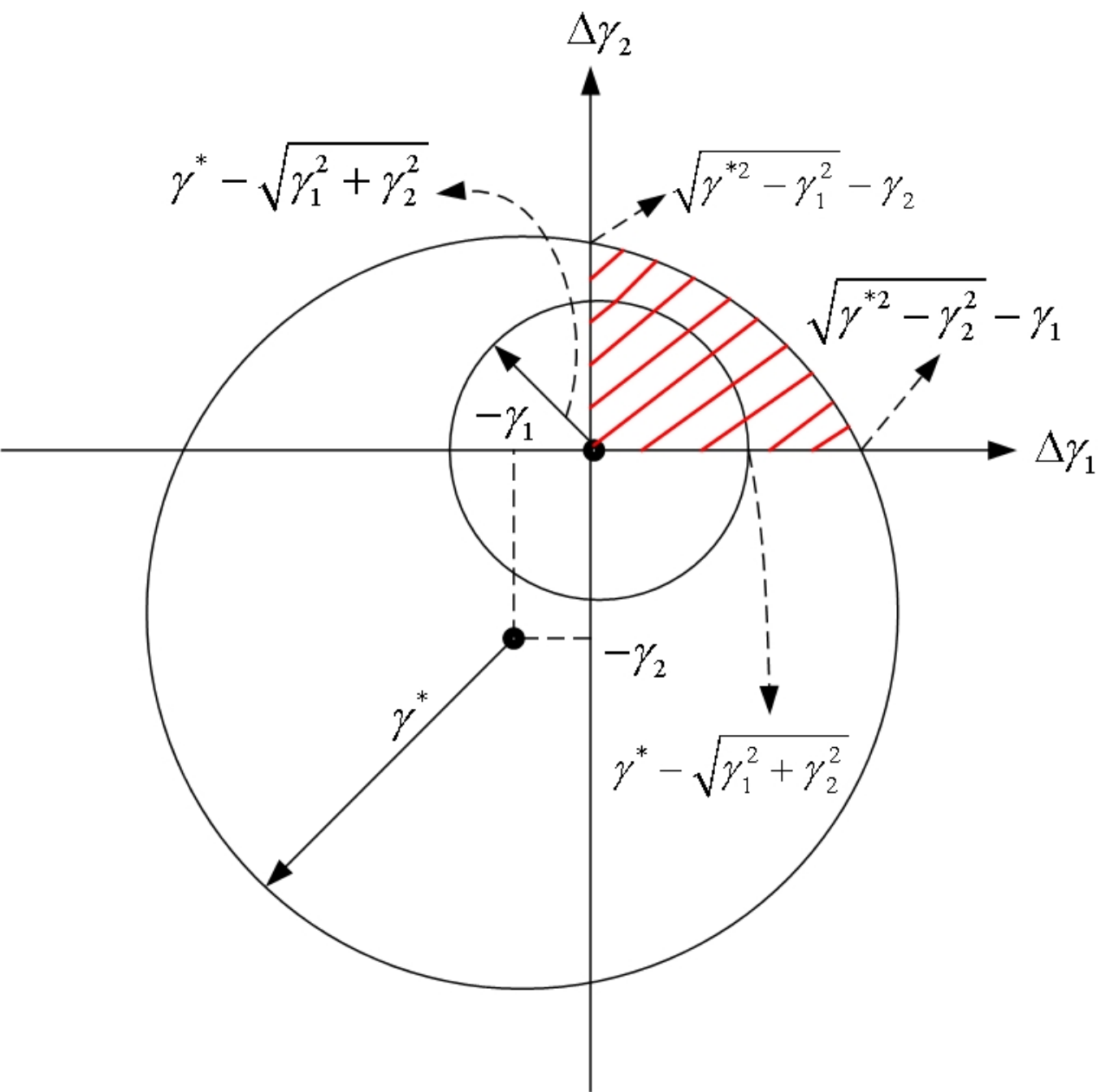}\\
  \caption{Geometric representation of uncertainty bounds. The admissible region is hachured.}\label{Fig1}
\end{figure}

\textbf{Remark 4.} Alternatively, we could write
\begin{eqnarray}
\begin{split}
\|\Omega_{\Delta}&(\xi_{1},u)-\Omega_{\Delta}(\xi_{2},u)\|\\
&\leq\|\Omega(\xi_{1},u)-\Omega(\xi_{2},u)\|+\|\Delta\Omega(\xi_{1},u)-\Delta\Omega(\xi_{2},u)\|\\
&\leq \sqrt{\gamma_{1}^{2}+\gamma_{2}^{2}}\|\xi_{1}-\xi_{2}\|+\sqrt{\Delta \gamma_{1}^{2}+\Delta \gamma_{2}^{2}}\|\xi_{1}-\xi_{2}\|\\
&=\left(\sqrt{\gamma_{1}^{2}+\gamma_{2}^{2}}+\sqrt{\Delta
\gamma_{1}^{2}+\Delta \gamma_{2}^{2}}\right)\|\xi_{1}-\xi_{2}\|.
\end{split}\notag
\end{eqnarray}
Then, we could conclude that, $\Delta\Phi(x,u)$ and $\Delta\Psi(x,u)$
can be any additive uncertainties with
\begin{align}
\sqrt{\Delta \gamma_{1}^{2}+\Delta \gamma_{2}^{2}}\leq
\gamma^{*}-\sqrt{\gamma_{1}^{2}+\gamma_{2}^{2}}.
\end{align}
However, it is not hard to show that
\begin{multline}
\sqrt{\left(\gamma_{1}+\Delta
\gamma_{1}\right)^{2}+\left(\gamma_{2}+\Delta \gamma_{2}\right)^{2}}
\leq \sqrt{\gamma_{1}^{2}+\gamma_{2}^{2}}+ \sqrt{\Delta \gamma_{1}^{2}+\Delta \gamma_{2}^{2}}\ , \ \ \
\forall \
\gamma_{1},\gamma_{2},\Delta\gamma_{1},\Delta\gamma_{2}\geq 0. \notag
\end{multline}
Therefore, the bound in \eqref{bound1} is less conservative. The
geometric representations of the two bounds are shown in Figure
\ref{Fig1}. The admissible region is hachured.

\section{Illustrative Example}
\begin{figure}[!h]
  \centering
  \includegraphics[width=.9\textwidth]{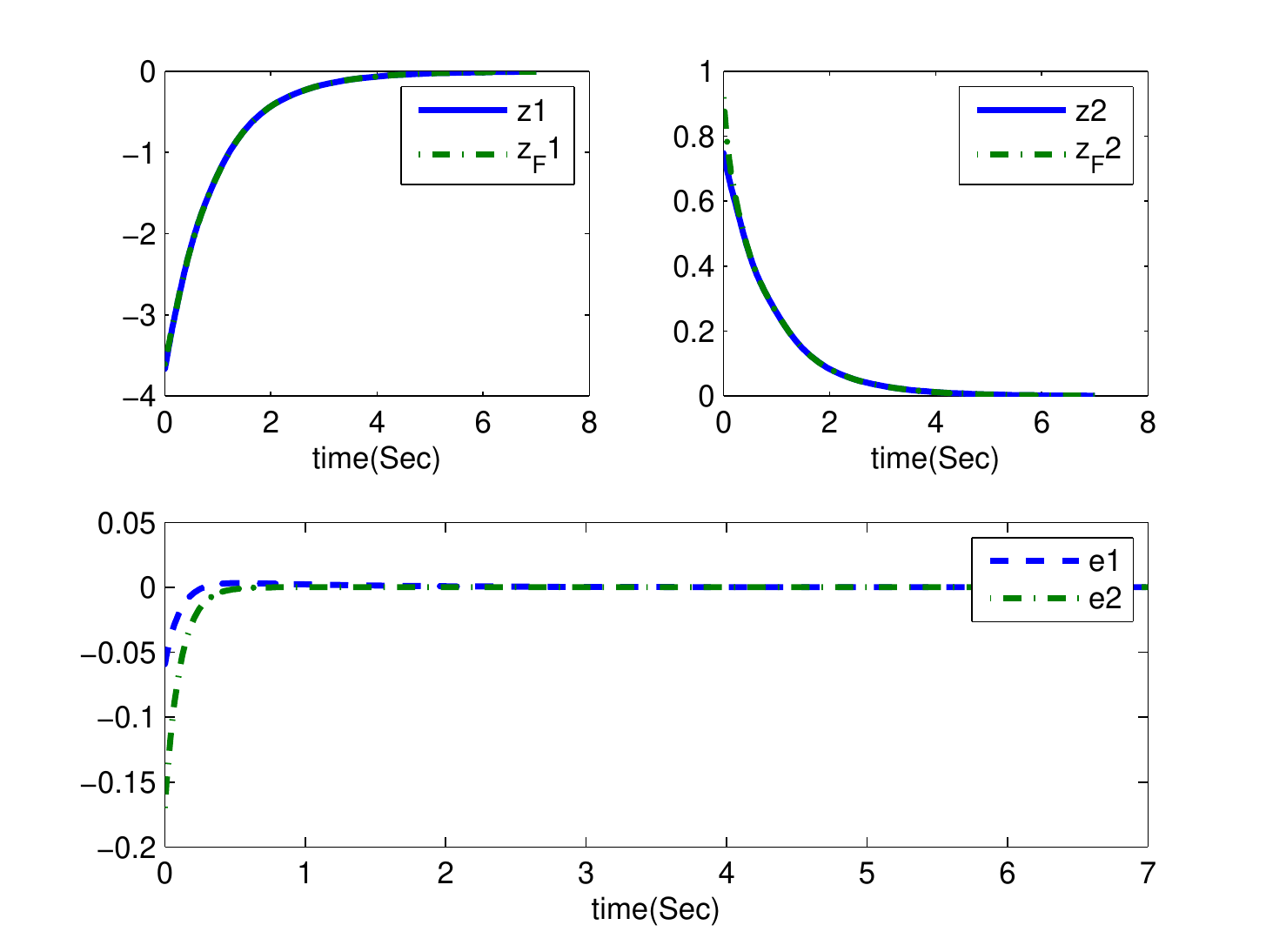}\\
  \caption{Simulations results of the nominal descriptor system and the $H_{\infty}$ filter}\label{Fig2}
\end{figure}

\begin{figure}[!h]
  \centering
  \includegraphics[width=.9\textwidth]{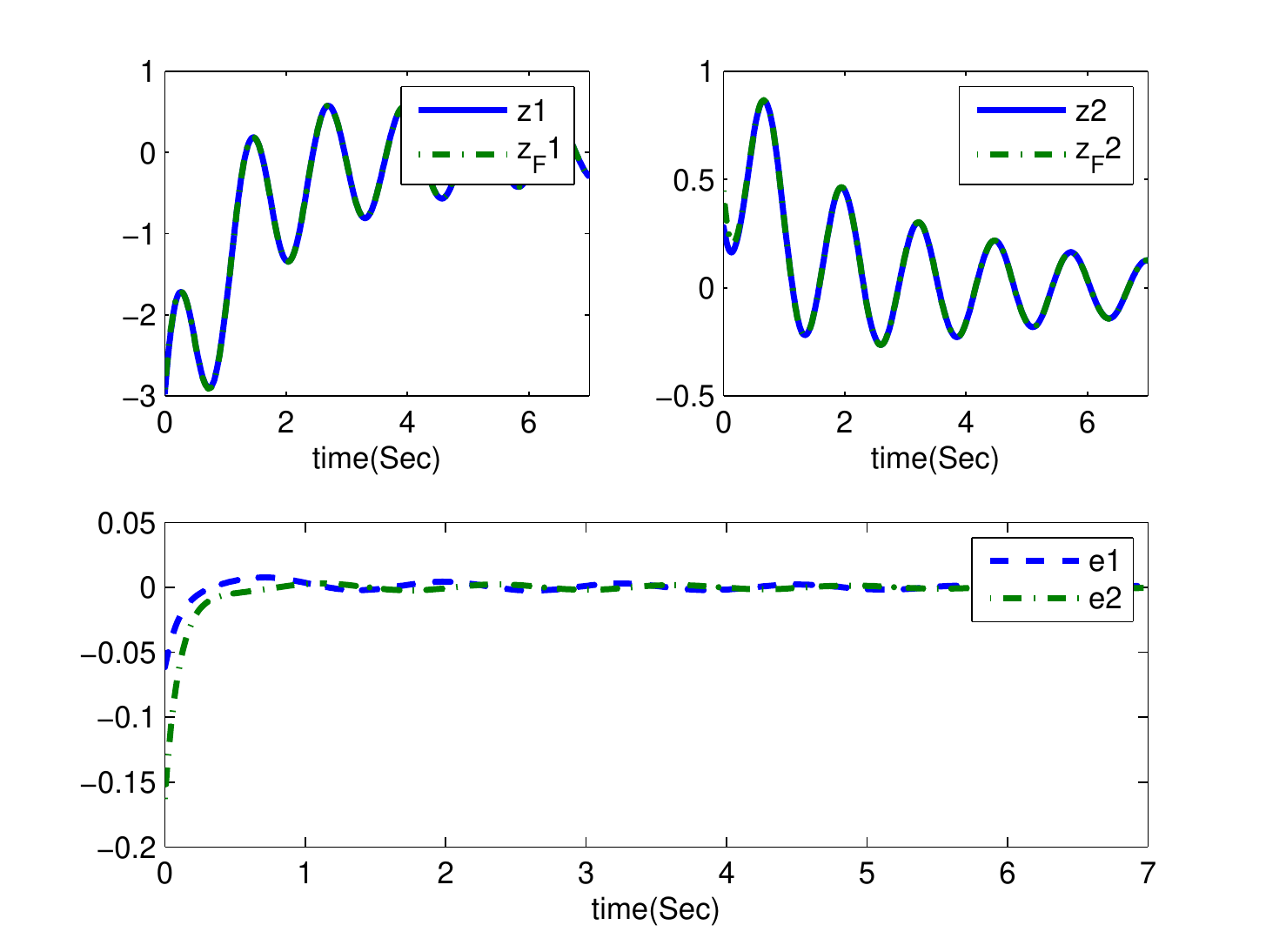}\\
  \caption{Simulations results of the descriptor system and the $H_{\infty}$ filter in the presence of disturbance}\label{Fig3}
\end{figure}

\begin{figure}[!h]
  \centering
  \includegraphics[trim= 30mm 65mm 30mm 65mm, clip, width=.9\textwidth]{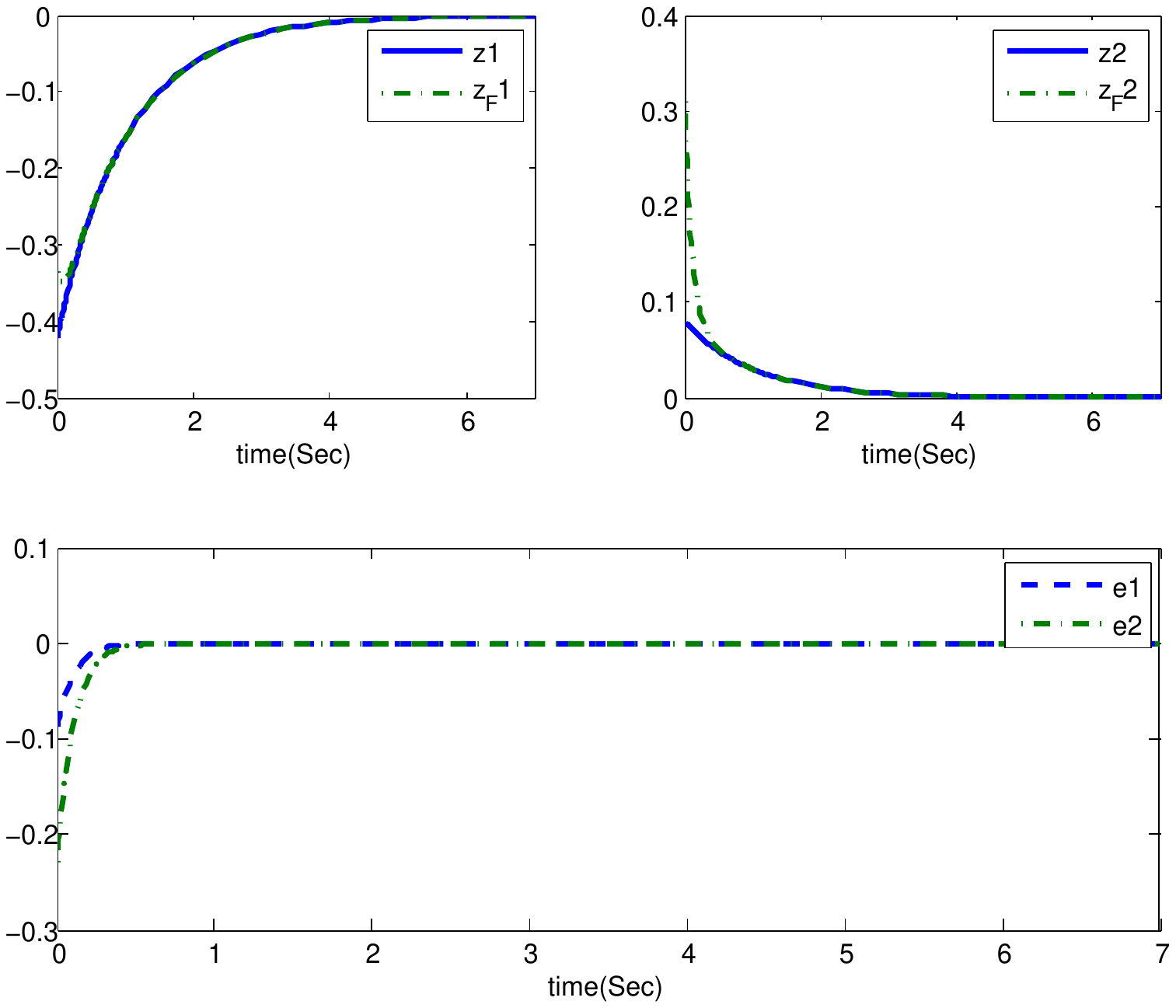}\\
  \caption{Simulations results of the descriptor system and the $H_{\infty}$ filter in the presence of model uncertainty}\label{Fig4}
\end{figure}

\begin{figure}[!h]
  \centering
  \includegraphics[trim= 30mm 65mm 30mm 65mm, clip, width=.9\textwidth]{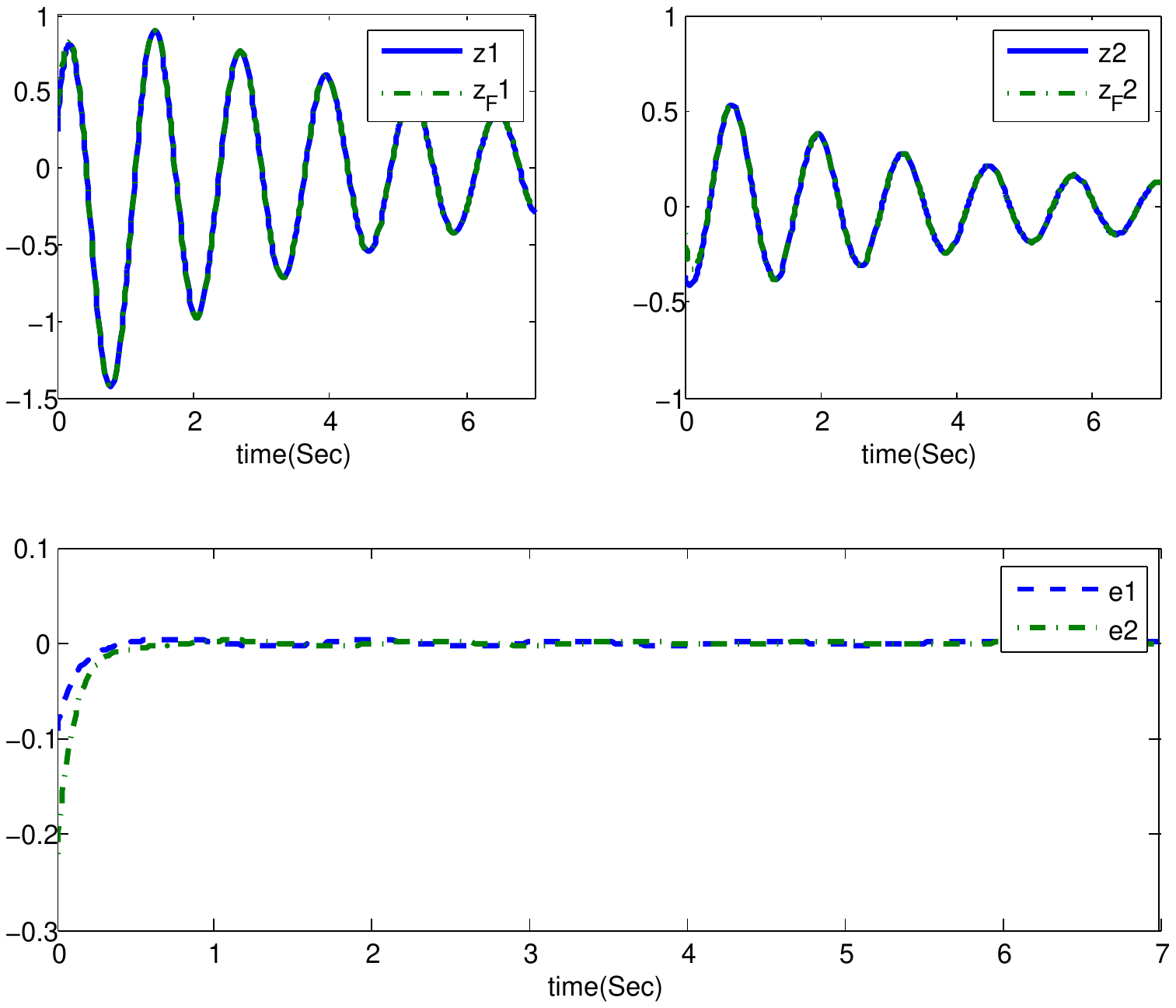}\\
  \caption{Simulations results of the descriptor system and the $H_{\infty}$ filter in the presence of model uncertainty and disturbance}\label{Fig5}
\end{figure}

Consider a system of class $\Sigma_{s}$ and suppose the nonimal system is given as
\begin{align}
&\left[
      \begin{array}{cc}
      2 & 3 \\
      4 & 6 \\
      \end{array}
\right]\left[
         \begin{array}{c}
           \dot{x}_{1} \\
           \dot{x}_{2} \\
         \end{array}
       \right]=\left[
    \begin{array}{cc}
      1 & 12 \\
      -6 & -15 \\
    \end{array}
  \right]\left[
         \begin{array}{c}
           x_{1} \\
           x_{2} \\
         \end{array}
       \right]+\frac{1}{2}\left[
        \begin{array}{c}
           \sin x_{2} \\
           \sin x_{1} \\
        \end{array}
      \right]\notag\\
&y= \left[
     \begin{array}{cc}
       1 & 0 \\
     \end{array}
   \right]\left[
         \begin{array}{c}
           x_{1} \\
           x_{2} \\
         \end{array}
       \right]\notag.
\end{align}
We assume the uncertainty and disturbances matrices as follows:
\begin{align}
M_{1}&=\left[
            \begin{array}{cc}
               0.1 & 0.1 \\
               -0.2 & 0.15 \\
            \end{array}
       \right], \ B=\left[
                          \begin{array}{c}
                            1 \\
                            1 \\
                          \end{array}
                     \right], \ N=\left[
                                     \begin{array}{cc}
                                        0.1 & 0 \\
                                        0 & 0.1 \\
                                     \end{array}
                                   \right]\notag\\
M_{2}&=\left[
         \begin{array}{cc}
           -0.25 & 0.25 \\
         \end{array}
       \right], \ \ D=0.2 \notag.
\end{align}
The system is globally Lipschitz with $\gamma=0.5$.
Now, we design a filter with dynamic structure. Therefore,
we have $\mathcal{E}_{1}=I$ and $\mathcal{E}_{2}=0$.
Using Corollary 1 with $\mu=0.25$ and $H=0.25I_{2}$, a robust
$H_{\infty}$ dynamic filter is obtained as:
\begin{align}
A_{F}&=\left[
        \begin{array}{cc}
        -70.9676 & -44.2097 \\
         12.2131 & -40.8541 \\
        \end{array}
      \right], \ B_{F}=\left[
         \begin{array}{c}
           4.3896 \\
           0.7843 \\
         \end{array}
       \right]\notag\\
C_{F}&=\left[
         \begin{array}{cc}
           -0.0057 &  -0.0043\\
           0.0047  & -0.0050 \\
         \end{array}
       \right], \ D_{F}= 1e-4 \times \left[
              \begin{array}{c}
                    -0.3877\\
                    0.1026 \\
              \end{array}
            \right]\notag\\
\epsilon_{1}&=1.2903 ,\ \ \ \epsilon_{2}=1.4497\notag\\
\alpha_{1}&=2.1406e-4, \ \ \ \alpha_{2}=1.0024\notag\\
\gamma^{*}&=0.9988\notag.
\end{align}
As mentioned earlier, in order to simulate the system,
we need consistent initial conditions. Matrix $\mathbf{E}$
is of rank $1$, thus, the system has $1$ differential equation and $1$
algebraic constraint. The system is currently in the implicit descriptor form.
In order to extract the algebraic constraint, we can convert the system
into semi-explicit differing algebraic. The matrix $\mathbf{E}$ can be decomposed as:
\begin{align}
\mathbf{E}=\left[
      \begin{array}{cc}
      2 & 3 \\
      4 & 6 \\
      \end{array}
\right]=S \left[
      \begin{array}{cc}
      1 & 0 \\
      0 & 0 \\
      \end{array}
\right]T,\notag
\end{align}
where
\begin{align}
S = \left[
      \begin{array}{cc}
      1 & 0 \\
      2 & 1 \\
      \end{array}
\right],\ T = \left[
      \begin{array}{cc}
      3 & 2 \\
      0 & 1 \\
      \end{array}
\right].\notag
\end{align}
Now, with the change of variables $\bar{x}=Tx$, the state equations in the original
system are rewritten in the \emph{semi-explicit} form as follows:
\begin{align}
\left[
      \begin{array}{cc}
      1 & 0 \\
      0 & 0 \\
      \end{array}
\right]\left[
         \begin{array}{c}
           \dot{\bar{x}}_{1} \\
           \dot{\bar{x}}_{2} \\
         \end{array}
       \right]=&\left[
    \begin{array}{cc}
      \frac{1}{3} & \frac{34}{3} \\
      -\frac{8}{3} & -\frac{101}{3} \\
    \end{array}
  \right]\left[
         \begin{array}{c}
           \bar{x}_{1} \\
           \bar{x}_{2} \\
         \end{array}
       \right]+\left[
       \begin{array}{cc}
      \frac{1}{2} & 0 \\
      -1 & \frac{1}{2} \\
    \end{array}
    \right]\left[
        \begin{array}{c}
           \sin \bar{x}_{2} \\
           \sin(\frac{1}{3}\bar{x}_{1}-\frac{2}{3}\bar{x}_{2}) \\
        \end{array}
      \right].\notag
\end{align}
So, the system is clearly decomposed into differential and algebraic parts.
The second equation in the above which is:
\begin{align}
-\frac{8}{3}\bar{x}_{1}-\frac{101}{3}\bar{x}_{2}-sin \bar{x}_{1}+\frac{1}{2}sin(\frac{1}{3}\bar{x}_{1}-\frac{2}{3}\bar{x}_{2})=0,\notag
\end{align}
is the algebraic equation which must be satisfied by the initial conditions. A set of consistent initial conditions satisfying the above equation is found as $\bar{x}_{1}(0)=-38.1034,\ \bar{x}_{2}(0)=3.0014$ which corresponds to $x_{1}(0)=-14.7020,\ x_{2}(0)=3.0014$ which in turn corresponds to $z_{1}(0)=-3.6755,z_{2}(0)=0.7503$, where $z=Hx$. Similarly, we find another set of consistent initial conditions for simulating the designed filter. Note that the introduced change of variables is for clarification purposes only to reveal the algebraic constraint which is \emph{implicit} in the original equations which facilitates calculation of consistent initial conditions, and is not required in the filter design algorithm. Consistent initial conditions could also be calculated using the original equations and in fact, most DAE solvers contain a built-in mechanism for consistent initialization using the descriptor form directly. Figure \ref{Fig2} shows the simulation results of $z$ and $z_{F}$ of the nominal system in the absence of disturbance and uncertainties,
where $z_{F}$ is the output of the filter as in \eqref{observer1}.
Now suppose an unknown $\mathcal{L}_{2}$ exogenous disturbance signal is affecting the system as $w(t)=50\exp(-0.2t)\cos(5t)$.
Figure \ref{Fig3} shows the simulation results of $z$ and $z_{F}$ in the presence of disturbance.
As expected, in the presence of disturbance, the observer filter error does not converge to zero (as long as the disturbance exists)
but it is kept in the vicinity of zero such that the norm bound $\|e\| \le \mu \|w\|$ is satisfied.
The designed filter guarantees $\mu$ to be at most $0.25$. The actual value of $\mu$ for this simulation is $0.0133$.
In the next step, we simulate the system in the presence of model uncertainty.
An (unknown to the filter) time-varying matrix considered is as $F(t)=diag\left(\frac{t}{t+0.1},\frac{t^2+0.1}{t^2+1}\right)$.
It is easy to verify that $F^{T}(t)F(t) \leq I$ for all $t$. Figure \ref{Fig4} shows the simulation results of $z$ and $z_{F}$ in the presence of model uncertainty.
As seen in the figure, the filter is robust against model uncertainty.
Finally, we simulate the system in the presence of both model uncertainty and disturbance. Figure \ref{Fig5} shows the simulation results.
The actual value of $\mu$ for this simulation is $0.0187$.

\section{Conclusions and Future Research Directions}
In this work, a new nonlinear $H_{\infty}$ dynamical filter design method for a
class of nonlinear descriptor uncertain systems is proposed through
semidefinite programming and strict LMI optimization. The developed
LMIs are linear both in the admissible Lipschitz constant and the
disturbance attenuation level allowing both two be an LMI optimization
variable. The proposed dynamical structure has more degree of freedom
than the conventional static-gain filters and is capable of robustly
stabilizing the filter error dynamics for some of those systems for which an
static-gain filter cannot be found. In addition, when the
static-gain filter also exists, the maximum admissible Lipschitz
constant obtained using the proposed dynamical filter structure
can be much larger than the at of the static-gain filter. The
achieved $H_{\infty}$ filter guarantees asymptotic stability of
the error dynamics and is robust against Lipschitz additive nonlinear
uncertainty as well as time-varying parametric uncertainty.

In the following we briefly discuss some future research avenues.

\begin{description}
                            \item[\emph{energy-to-peak filtering}:] As mentioned in the Introduction, in $H_{\infty}$ filtering, the $\mathcal{L}_{2}$ gain from the exogenous disturbance to the filter error is guaranteed
to be less than a prespecified level, making the underlying $\mathcal{L}_{2}$ gain minimization an energy-to-energy performance criterion. An alternative approach is the so-called $\mathcal{L}_{2}-\mathcal{L}_{\infty}$ filtering. In $\mathcal{L}_{2}-\mathcal{L}_{\infty}$ filtering,
the ratio of the peak value of the error ($\mathcal{L}_{\infty}$ norm)
to the energy of disturbance ($\mathcal{L}_{2}$ norm) is considered, therefore,
conforming an \emph{energy-to-peak} performance criterion. The tools and methods provided in this work can be used to solve the robust $\mathcal{L}_{2}-\mathcal{L}_{\infty}$ filtering
problem for the studied class of nonlinear descriptor systems.
                            \item[\emph{mixed $H_{2}/H_{\infty}$ filtering}:] One the main advantages of $H_{\infty}$ approach is that it does not require any knowledge about
the statistics of noise. It works for every finite energy signal.
The noise terms may be random with possibly unknown statistics, or they may be deterministic.
If the statistics of noise are known, Kalman filtering (i.e. $H_{2}$) approaches can be used. However, estimating the statistics of noise is a difficult task and
the representation of disturbances by white noise processes are often unrealistic. That was one the main motives that $H_{\infty}$ approaches
developed at the first place. Nevertheless, the prior knowledge about the statistics of disturbance can be utilized to set up a mixed $H_{2}/H_{\infty}$
performance criterion.
                            \item[\emph{uncertainty in the mass matrix}:] To the best of the author's knowledge,
in all works on nonlinear uncertain descriptor systems (including this work), the mass matrix $\mathbf{E}$ is assumed to be fully known.
This is required because the matrix $\mathbf{E}$ participates in the construction of the generalized Lyapunov function. For models with unstructured uncertainty,
this is not a big deal. However, for models with structured parametric uncertainty, there might be cases that an intrinsic uncertainty is associated
with the elements of $\mathbf{E}$ (e.g. due to presence of uncertain parameters in $\mathbf{E}$), which cannot be incorporated into $A$ and $C$.
Therefore, considering a $\Delta \mathbf{E}$ might become inevitable. This is currently an open problem.
                          \end{description}


\bibliographystyle{plain}
\bibliography{References}

\begin{thebibliography}{10}

\bibitem{Abbaszadeh3}
Masoud Abbaszadeh and Horacio~J. Marquez.
\newblock Robust {$H_{\infty}$} observer design for a class of nonlinear
  uncertain systems via convex optimization.
\newblock {\em Proceedings of the 2007 American Control Conference, New York,
  U.S.A.}, pages 1699--1704.

\bibitem{abbaszadeh2008lmi}
Masoud Abbaszadeh and Horacio~J Marquez.
\newblock {LMI} optimization approach to robust {$H_{\infty}$} filtering for
  discrete-time nonlinear uncertain systems.
\newblock In {\em American Control Conference, 2008}, pages 1905--1910. IEEE,
  2008.

\bibitem{Abbaszadeh4}
Masoud Abbaszadeh and Horacio~J. Marquez.
\newblock Robust {$H_{\infty}$} observer design for sampled-data {Lipschitz}
  nonlinear systems with exact and {Euler} approximate models.
\newblock {\em Automatica}, 44(3):799--806, 2008.

\bibitem{Abbaszadeh5}
Masoud Abbaszadeh and Horacio~J. Marquez.
\newblock {LMI} optimization approach to robust {$H_{\infty}$} observer design
  and static output feedback stabilization for discrete-time nonlinear
  uncertain systems.
\newblock {\em International Journal of Robust and Nonlinear Control},
  19(3):313--340, 2009.

\bibitem{abbaszadeh2012generalized}
Masoud Abbaszadeh and Horacio~J Marquez.
\newblock A generalized framework for robust nonlinear {$H_{\infty}$} filtering
  of lipschitz descriptor systems with parametric and nonlinear uncertainties.
\newblock {\em Automatica}, 48(5):894--900, 2012.

\bibitem{Agamennonia.etal2004}
O.~Agamennonia, I.~Skrjanc, M.~Lepetic, H.~Chiacchiarinic, and D.~Matko.
\newblock Nonlinear uncertainty model of a magnetic suspension system.
\newblock {\em Mathematical and Computer Modelling}, 40(9-10):1075--1087, 2007.

\bibitem{Boulkroune.etal2010_2}
B.~Boulkroune, M.~Darouach, and M.~Zasadzinski.
\newblock Moving horizon state estimation for linear discrete-time singular
  systems.
\newblock {\em Control Theory Applications, IET}, 4(3):339 --350, march 2010.

\bibitem{Boulkroune.etal2010_1}
B.~Boulkroune and A.~Zemouche.
\newblock Robust fault diagnosis for a class of nonlinear descriptor systems.
\newblock In {\em Control and Fault-Tolerant Systems (SysTol), 2010 Conference
  on}, pages 335 --340, oct. 2010.

\bibitem{Boutayeb1}
M.~Boutayeb and M.~Darouach.
\newblock Observers design for nonlinear descriptor systems.
\newblock {\em Proceedings of the IEEE Conference on Decision and Control},
  3:2369--2374, 1995.

\bibitem{Cellier2006}
F.E. Cellier and E.~Kofman.
\newblock {\em {Continuous System Simulation}}.
\newblock Springer, 2006.

\bibitem{Dai}
L.~Dai.
\newblock {\em Singular control systems}, volume 118 of {\em Lecture Notes on
  Control and Information Sciences}.
\newblock Sprinter, 1989.

\bibitem{Darouach.etal2008_1}
M.~Darouach and L.~Boutat-Baddas.
\newblock Observers for lipschitz nonlinear descriptor systems: Application to
  unknown inputs systems.
\newblock In {\em Control and Automation, 2008 16th Mediterranean Conference
  on}, pages 1369 --1374, june 2008.

\bibitem{Darouch1}
M.~Darouach and M.~Boutayeb.
\newblock Design of observers for descriptor systems.
\newblock {\em IEEE Transactions on Automatic Control}, 40(7):1323--1327, 1995.

\bibitem{Darouch2}
M.~Darouach, M.~Zasadzinski, and M.~Hayar.
\newblock Reduced-order observer design for descriptor systems with unknown
  inputs.
\newblock {\em IEEE Transactions on Automatic Control}, 41(7):1068--1072, 1996.

\bibitem{deSouza1}
Carlos~E. de~Souza, Lihua Xie, and Youyi Wang.
\newblock {$H_{\infty}$} filtering for a class of uncertain nonlinear systems.
\newblock {\em Systems and Control Letters}, 20(6):419--426, 1993.

\bibitem{Fritzson2004}
Peter Fritzson.
\newblock {\em Principles of Object-Oriented Modeling and Simulation with
  Modelica 2.1}.
\newblock Wiley-IEEE Press, 2004.

\bibitem{Horn1}
R.~A. Horn and C.~R. Johnson.
\newblock {\em Matrix Analysis}.
\newblock Cambrige University Press, 1985.

\bibitem{Hou}
M.~Hou and P.~C. Muller.
\newblock Observer design for descriptor systems.
\newblock {\em IEEE Transactions on Automatic Control}, 44(1):164--169, 1999.

\bibitem{Ishihara}
J.~Y. Ishihara and M.~H. Terra.
\newblock On the {Lyapunov} theorem for singular systems.
\newblock {\em IEEE Transactions on Automatic Control}, 47(11):1926--1930,
  2002.

\bibitem{Khargonekar}
Pramod~P. Khargonekar, Ian~R. Petersen, and Kemin Zhou.
\newblock Robust stabilization of uncertain linear systems: Quadratic
  stabilizability and {$H_{\infty}$} control theory.
\newblock {\em IEEE Transactions on Automatic Control}, 35(3):356--361, 1990.

\bibitem{Kosut1997}
R.L. Kosut.
\newblock Nonlinear uncertainty model unfalsification.
\newblock In {\em American Control Conference, 1997. Proceedings of the 1997},
  volume~3, pages 2098--2102, 1997.

\bibitem{YALMIP}
J.~Lofberg.
\newblock {YALMIP} : A toolbox for modeling and optimization in {MATLAB}.
\newblock 2004.

\bibitem{Lu_G3}
G.~Lu and D.~W.~C. Ho.
\newblock Full-order and reduced-order observers for {Lipschitz} descriptor
  systems: The unified {LMI} approach.
\newblock {\em IEEE Transactions on Circuits and Systems II: Express Briefs},
  53(7), 2006.

\bibitem{Luck.etal2004}
R.~Luck and J.~W. Stevens.
\newblock A simple numerical procedure for estimating nonlinear uncertainty
  propagation.
\newblock {\em ISA Transations}, 43(4):491--497, 2004.

\bibitem{Marquez}
H.~J. Marquez.
\newblock {\em Nonlinear Control Systems: Analysis and Design}.
\newblock Wiley, NY, 2003.

\bibitem{Masubuchi}
I.~Masubuchi, Y.~Kamitane, A.~Ohara, and N.~Suda.
\newblock {$H_{\infty}$} control for descriptor systems: A matrix inequalities
  approach.
\newblock {\em Automatica}, 33(4):669--673, 1997.

\bibitem{Pantelides}
Constantinos~C. Pantelides.
\newblock The consistent initialization of differential-algebraic systems.
\newblock {\em SIAM Journal on Scientific Computing}, 9(2):219--231, 1988.

\bibitem{Rehman.etal2011}
Obaid~Ur Rehman, Baris Fidan, and Ian~R. Petersen.
\newblock Uncertainty modeling and robust minimax {LQR} control of
  multivariable nonlinear systems with application to hypersonic flight.
\newblock {\em Asian Journal of Control}, 2011.

\bibitem{McKenna.etal2004}
McKenna~L. Robertsa, James~W. Stevensa, , and Rogelio Luck.
\newblock Evaluation of parameter effects in estimating non-linear uncertainty
  propagation.
\newblock {\em Measurement}, 40(1):15--20, 2007.

\bibitem{Boyd}
E.~Feron S.~Boyd, L. El~Ghaoui and V.~Balakrishnan.
\newblock {\em Linear matrix inequalities in system and control theory}.
\newblock SIAM, PA, 1994.

\bibitem{Shields}
D.~N. Shields.
\newblock Observer design and detection for nonlinear descriptor systems.
\newblock {\em International Journal of Control}, 67(2):153--168, 1997.

\bibitem{SeDuMi}
Jos~F. Sturm.
\newblock {SeDuMi}.
\newblock 2001.

\bibitem{Uezato}
Eiho Uezato and Masao Ikeda.
\newblock Strict {LMI} conditions for stability, robust stabilization, and
  {$H_{\infty}$} control of descriptor systems.
\newblock {\em Proceedings of the 38th IEEE Conference on Decision and
  Control}, 4:4092--4097, 1999.

\bibitem{He_Wang}
He-Sheng Wang, Chee-Fai Yung, and Fen-Ren Chang.
\newblock {\em {$H_{\infty}$} control for Nonlinear Descriptor Systems}, volume
  326 of {\em Lecture Notes in Control and Information Sciences}.
\newblock Springer, 2006.

\bibitem{deSouza2}
Youyi Wang, Lihua Xie, and Carlos~E. de~Souza.
\newblock Robust control of a class of uncertain nonlinear systems.
\newblock {\em Systems and Control Letters}, 19(2):139--149, 1992.

\bibitem{Zimmer}
G.~Zimmer and J.~Meier.
\newblock On observing nonlinear descriptor systems.
\newblock {\em Systems and Control Letters}, 32(1):43--48, 1997.

\end{thebibliography}


\end{document}